\documentclass[aps,prd,twocolumn,showpacs,groupedaddress,nofootinbib]{revtex4}

\usepackage{graphicx}
\usepackage{longtable}

\newcommand {\Dtau}{\Delta\tau}

\begin{document}

\title{\bf
PATH INTEGRAL MONTE CARLO APPROACH TO THE U(1) LATTICE GAUGE THEORY IN (2+1) 
DIMENSIONS}

\author{Mushtaq Loan}
\email{mushe@phys.unsw.edu.au}
\author{Michael Brunner}
\author{Clare Sloggett}
\author{Chris Hamer}

\affiliation{School of Physics, The University of New South Wales,
Sydney, NSW 2052, Australia}

\date{September 20, 2002}

\begin{abstract}
 
 Path Integral Monte Carlo simulations have been performed for U(1) 
lattice gauge theory in (2+1) dimensions on anisotropic lattices. We extract 
the static quark potential, the string tension and the low-lying ``glueball" 
spectrum. 
The Euclidean string tension and mass gap decrease exponentially at weak
coupling in excellent agreement with the predictions of Polyakov and
G{\" o}pfert and Mack, but their magnitudes are five times bigger than
predicted.
Extrapolations are made to the extreme anisotropic 
or Hamiltonian
limit, and comparisons are made with previous estimates obtained in the 
Hamiltonian formulation.
\end{abstract}


\maketitle

\section{Introduction}

Classical Monte Carlo simulations\cite{cre79} of the path integral in 
Euclidean lattice 
gauge theory\cite{wil74} have been very successful, and this is currently 
the preferred method
for {\it ab initio} calculations in quantum chromodynamics (QCD) in the low energy 
regime. Monte Carlo approaches to the Hamiltonian version of QCD 
propounded by
Kogut and Susskind\cite{kog75} have been less successful, however, and lag at 
least ten years
behind the Euclidean calculations. Our aim in this paper is to see whether 
useful
results can be obtained for the Hamiltonian version by using the standard 
Euclidean
Monte Carlo methods for anisotropic lattices\cite{mor97}, and extrapolating 
to the 
Hamiltonian limit in which the time variable becomes continuous, i.e. the
lattice spacing in the
time direction goes to zero. The Hamiltonian version of lattice gauge theory 
is less popular than the Euclidean version, but is still worthy of study. It
can provide a valuable check of the universality of the Euclidean 
results\cite{ham96},
and it allows the application of many techniques imported from 
quantum many-body
theory and condensed matter physics, such as strong coupling 
expansions\cite{ban77a},
the t-expansion\cite{hor84}, the coupled-cluster method\cite{guo88},
 the plaquette expansion \cite{john97}, loop representation method \cite{aro93}, 
 and more recently the density matrix renormalization group\cite{byr02}
(DMRG). None of these
techniques has proved as useful as Monte Carlo in (3+1) dimensions; but in 
lower 
dimensions they are more competitive.

A number of Quantum Monte Carlo methods have been applied to Hamiltonian
lattice gauge theory in the past, with somewhat mixed results. 
A ``Projector Monte Carlo"
approach\cite{bla83,deg85} using a strong coupling representation for the gauge 
fields runs into
difficulties for non-Abelian models, in that it requires Clebsch-Gordan 
coefficients
for SU(3) which are not even known at high orders; and furthermore a version
of the ``minus sign problem" rears its head\cite{ham94}.
A Greens Function Monte Carlo approach was pioneered by Chin et al\cite
{chi84}
and
Heys and Stump\cite{hey83}, which uses a weak coupling representation for the 
gauge fields.
This approach can be used successfully for  non-Abelian theories, and obtains
estimates of parameters such as the string tension and glueball masses from the
correlation functions in a similar fashion to Euclidean techniques.
Unfortunately the approach requires the use of a ``trial wave function" to
guide random walkers in the ensemble towards the preferred regions of
configuration space\cite{kal66}. This introduces a variational element into the
procedure, in that the results may exhibit a systematic dependence on the trial
wave function. We have previously explored\cite{sam99,ham00,ham00a} 
a ``forward-walking"
technique\cite{liu74,whi79}
for measuring the expectation values and the correlation functions,
which should minimize this dependence; but calculations for the SU(3) 
Yang-Mills
theory in (3+1) dimensions still showed an unacceptably strong sensitivity to 
the
parameters of the trial wave function\cite{ham00a}. 
For this reason, we are forced 
to look
yet again for an alternative approach.

 As mentioned above our aim in 
this paper is to use
standard Euclidean path integral Monte Carlo techniques for anisotropic 
lattices, and
see whether useful results can be obtained in the Hamiltonian limit.
Morningstar and Peardon\cite{mor97} showed some time ago that the use of
anisotropic lattices can be advantageous in any case, particularly for
the measurement of glueball masses. We use a number of their techniques
in what follows. 

As a first trial of this approach, we treat the U(1) gauge model in
(2+1)D, which is one of the simplest models with dynamical gauge degrees
of freedom, and has also been studied extensively by other means (see
Section II). Path integral Monte Carlo methods were applied to this
model a long time ago by Hey and collaborators\cite{amb82,cod86}, but
the techniques used at that time were not very sophisticated, and the
results were rather qualitative. Very little has been done since then
using this approach on this particular model, apart from a calculation by 
Irb{\" a}ck and Peterson \cite{irb87}.  

The rest of this paper is organised as follows. In section II we discuss 
the U(1) model in (2+1) dimensions in its lattice formulation, and
outline some of the work done on it previously. The details 
of the simulations, including the generation of the gauge-field
configurations, the 
construction of the Wilson loop operators and glueball operators, and the 
extraction of the potential and string tension  
estimates are described in section III.  
In section IV  we present 
our main results for the mean plaquette, static quark potential, string 
tension and glueball masses.  The static
quark potential has not previously been exhibited for this model, as far
as we are aware. Finally 
we make an extrapolation to the Hamiltonian limit, and comparisons are 
made with estimates obtained by other means in that limit. 
We find that indeed the PIMC method can give better results than other
weak-coupling Monte Carlo methods, even in the Hamiltonian limit.
Our conclusions are summarized in Sec. VI.  

\section{THE U(1) MODEL}

Consider the  isotropic Abelian U(1) lattice gauge theory in three dimensions.
 The theory is
defined by the action\cite{wil74} 
\begin{equation}
S = \beta 
\sum_{r,\mu,\nu}\mbox{ReTr}P_{\mu\nu}
\label{eqn1}
\end{equation}
where
\begin{equation}
\mbox{P}_{\mu\nu}(r) =\left[1-
Re Tr\{\mbox{U}_{\mu}(r)\mbox{U}_{\nu}(r+\hat{\mu})
\mbox{U}_{\mu}^{\dagger}(r+\hat{\nu})\mbox{U}_{\nu}^{\dagger}(r)\right\}]
\label{eqn2}
\end{equation}
is the plaquette variable given by the product of the link variables taken around an
elementary plaquette. The link variable  $\mbox{U}_{\mu}(r)$ is defined by
 \begin{equation}
 \mbox{U}_{\mu}(r) =\mbox{exp}[ieaA_{\mu}(r)] = 
\mbox{exp}[i\theta_{\mu}(r)]
\label{eqn3}
 \end{equation}
 where in the compact form of the model, $\theta_{\mu}(r) 
 (=eaA_{\mu}(r)) \in [0,2\pi]$
 represents the gauge field  on the directed link $r \rightarrow r+\hat{\mu}$.
 The parameter $\beta$ is related to the bare gauge coupling by
 \begin{equation}
 \beta =\frac{1}{g^{2}}
\label{eqn4}
 \end{equation}
 where $g^{2} = ae^{2}$, in (2+1) dimensions.

The lattice U(1) model in (2+1) dimensions has been studied by many authors, 
and possesses some important
similarities with QCD (for a more extensive review, see for example ref.
\cite{ham94}). If one takes the ``naive" continuum limit at a fixed
energy scale, one regains the simple continuum theory of non-interacting
photons\cite{gro83}; but if one renormalizes or rescales in the standard
way so as to maintain the
mass gap constant, then one obtains a confining theory of free massive bosons. 
Polyakov\cite{pol78} showed that a linear potential appears between
static charges due to instantons in the lattice theory; and
G{\" o}pfert and Mack\cite{gop82} proved that in the continuum limit the
theory converges to a scalar free field theory of massive bosons. 
They found that in that limit the mass gap behaves as 
\begin{equation}
am_D = \sqrt{\frac{8\pi^2}{g^2}}\exp(-\frac{\pi^2}{g^2}v(0))
\label{eqn5}
\end{equation}
while the string tension is bounded by
\begin{equation}
a^2\sigma \geq c\sqrt{\frac{g^2}{2\pi^2}}\exp(-\frac{\pi^2}{g^2}v(0))
\label{eqn6}
\end{equation}
where $v(0)$ is the Coulomb potential at zero distance, and has a value
in lattice units
\begin{equation}
v(0) = 0.2527
\label{eqn7}
\end{equation}
for the isotropic case. They argue that (\ref{eqn6}) represents the true
asymptotic behaviour of the string tension, where the constant $c$ is
equal to $8$ in classical approximation.
The theory has a non-vanishing string tension for arbitrarily large
$\beta$, similar to the behaviour expected for non-Abelian lattice gauge
theories in four dimensions.

 For an anisotropic lattice, the gauge action becomes\cite{mor97} 
 \begin{equation}
 S = \beta_{s}\sum_{r,i<j}P_{ij}(r) +\beta_{t}\sum_{r,i}P_{it}(r)
\label{eqn8}
 \end{equation}
 where
  $P_{ij}$ and $P_{it}$ are the spatial and temporal plaquette 
variables respectively.
 In the classical limit
 \begin{eqnarray}
 \beta_{s} & = & \frac{a_{t}}{e^{2}a_{s}^{2}}  = \frac{1}{g^{2}}\Dtau \\
 \beta_{t} & = & \frac{1}{e^{2}a_{t}}  = \frac{1}{g^{2}}\frac{1}{\Dtau}
 \label{eqn10}
 \end{eqnarray}
 where $\Dtau = a_{t}/a_{s} $ is the anisotropy 
parameter, $a_{s}$ is the lattice
 spacing in the space direction, and $a_{t}$ 
is the temporal spacing.
 The above action can be written as
 \begin{eqnarray}
 S & = & \beta \left[\Dtau \sum_{r}\sum_{i<j}\bigg(1-\cos \theta_{ij}(r)\bigg)
\right.
\nonumber\\
& & \left.+\frac{1}{\Dtau }\sum_{r,i}\bigg(1-\cos \theta_{it}(r)\bigg)
\right]
\label{eqn11}
\end{eqnarray} 

In the limit $\Delta \tau
 \rightarrow 0$, the time variable becomes 
continuous, and
 we obtain the Hamiltonian limit of the model 
(modulo a Wick rotation back to Minkowski space).

The behaviour of the mass gap in the anisotropic case will be similar
to equation (\ref{eqn5}) Generalizing
discussions by Banks {\it et al}\cite{ban77} and
Ben-Menahem\cite{ben79}, we find that the exponential
factor takes exactly the same form in the anisotropic case.
The only difference is that the lattice Coulomb potential at zero
spacing for general $\Delta \tau$ is
\begin{eqnarray}
v(0) & = & \int^{\pi}_{-\pi} \frac{d^{3}k}{(2\pi)^{3}} \frac{\Delta
\tau}{4[\sin^{2}(k_0/2) + \Delta \tau^{2}(\sin^{2}(k_1/2) +
\sin^{2}(k_2/2))]} \nonumber\\
 & = & \left\{ \begin{array} {c}
 0.2527 \hspace{5mm} (\Delta \tau = 1) \\
 0.3214 \hspace{5mm} (\Delta \tau = 0)
\end{array} 
\right.
\label{eqn13}
\end{eqnarray}
But this result neglects the effects of monopoles with charges  other
than $ 0, \pm 1$ in the monopole gas, which is justified in the
Euclidean case, but not in the Hamiltonian limit\cite{ban77,ben79}.

The Hamiltonian version of the model has been studied by many methods:
some recent studies include series expansions
\cite{ham92}, finite-lattice techniques\cite{irv83}, the t-expansion
\cite{hor87,mor92}, and coupled-cluster techniques
\cite{dab91,fan96,bak96} and the plaquette expansions \cite{john97}
 as well
as Quantum Monte Carlo 
methods \cite{chi84,koo86,yun86,ham94,ham00}. 
Quite accurate estimates have been obtained for the string
tension and the mass gaps, which can be used as comparison for our 
present results. The finite-size scaling
properties of the model can be predicted using an effective Lagrangian 
approach combined with a weak-coupling expansion
\cite{ham93}, and the predictions agree very well with finite-lattice data
\cite{ham94}.

\section{METHODS}

\subsection{Path Integral Monte Carlo algorithm}

We perform 
standard path integral
Monte Carlo simulations on a finite lattice of size $N_{s}^{2}\times 
N_{\tau}$, where
$N_{s}$ is the number of lattice sites in the space direction and 
$N_{\tau}$ 
in the temporal direction, with spacing ratio 
$\Delta \tau = a_{t}/a_{s} $.
By varying $\Delta \tau $ it is possible to change $a_{t}$, while 
keeping the spacing in the spatial direction fixed.  
The
simulations were performed on lattices with $N_s=16$ sites in each of 
the two
spatial directions and  
$N_{t}= 16 - 64$ in the temporal direction for a range
of couplings $\beta = 1 - 3$.

The ensembles of field configurations were generated by using a Metropolis 
algorithm. Starting from an arbitrary initial configuration of link angles, we 
successively update
link angles $\theta _{\mu}(\vec{r},\tau)$
at positions $(\vec{r},\tau)$ which are chosen randomly each time. 
We propose a  change $\Delta\theta$
to  this link angle, which is randomly drawn from a uniform 
distribution on $[-\Delta,\Delta]$, where $\Delta$ is adjusted for
each set of parameters to give an acceptable ``hit rate" around 70-80\%.
The change is accepted or rejected 
according to the standard Metropolis procedure.

For high anisotropy ($\Delta \tau <<1$), any change in a 
time-like plaquette will produce a large change in the action, 
whereas changes to the 
space-like plaquettes will cause a much smaller change in the action. 
This makes the system very ``stiff" against variations in the time-like 
plaquettes, and therefore very slow to equilibrate, with long 
autocorrelation times.
To alleviate this problem, we used a Fourier update 
procedure\cite{bat85,dav90}. Here proposed changes are made to 
space-like links which 
are designed to alter space-like plaquette values much more than the 
time-like plaquette values. At randomly chosen intervals and random
locations,  
we propose a non-local change 
$\Delta\theta (\vec{r},\tau)= X\sin k(\tau -\tau_{0} )$ on a ``ladder" 
of 
space-like links extending half a wavelength $(\lambda = 2\pi /k)$ in the 
time direction, where both $k$ and $X$ are randomly chosen at each 
update from uniform 
distributions in suitably chosen intervals.
We replaced approximately $30\%$ of the ordinary 
Metropolis updates with Fourier updates for anisotropy $\Delta \tau 
>0.444 $ and $50\%$ for highly anisotropic cases ($\Delta \tau
<0.444 $). These moves satisfy the requirements of detailed balance and
ergodicity for the algorithm.

\begin{figure}[!h]
\scalebox{0.45}{\includegraphics{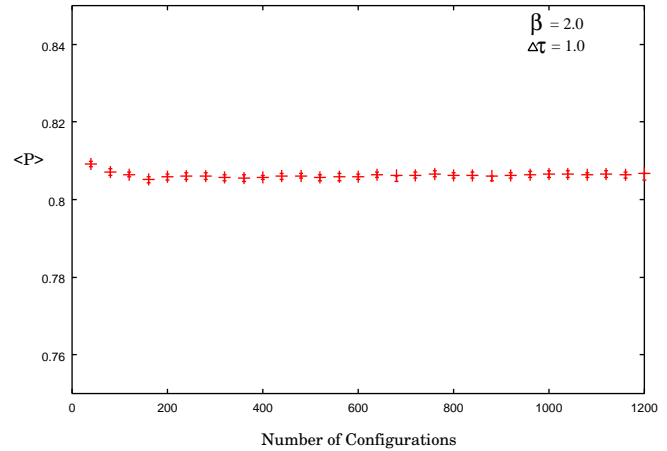}}
\caption{
\label{fig1}
Plot of the mean plaquette value against the number of 
configurations.}
\end{figure}

A single 
sweep involves attempting $N$ changes to randomly chosen links of the 
lattice, where $N (=3N_{t}N_s^{2})$ is the total number of links on the 
lattice. The first several thousand sweeps are discarded 
to allow the system to 
relax to equilibrium.
Figure \ref{fig1}  
shows measurements of the mean plaquette  
for $\beta = 2.0$ and $\Delta \tau =1.0 $, and it can be seen that 
equilibrium is reached after about 50,000 sweeps,
with the
measurements fluctuating about the equilibrum value thereafter.
For highly anisotropic cases the system was much slower to equilibrate,
despite the Fourier acceleration, and in the worst case the
equilibration time was of order 100,000 sweeps.

After discarding the initial sweeps, the 
configurations were stored every 250 sweeps
thereafter  for later analysis. 
Ensembles of about 1,000 configurations were stored  to 
measure the static quark potential and glueball masses
 at each $\beta$ for $\Delta \tau \geq 0.4$,
and 1,400 configurations for $\Delta \tau \leq 0.333$.
 Measurements made on
these stored configurations were grouped into 5 blocks, and then the
mean and standard deviation of the final quantities were estimated by
averaging over the `block averages', treated as independent
measurements. Each block average thus comprised 50,000 - 70,000 sweeps.

\subsection{Interquark Potential}

The static quark-antiquark potential, $V(\bf {r})$ for various spatial 
separations $\bf{r}$ is extracted from the expectation values of the 
Wilson loops. 
 The timelike Wilson 
loops  are expected to behave as:
 \begin{equation}
W({\bf r},\tau) \simeq Z({\bf r})\mbox{exp}\big[-\tau V({\bf r})] + 
\mbox{(excited state contributions)}
\label{eqn14}
\end{equation}
We have averaged only over loops $({\bf x_{0}},\tau_{0}) \rightarrow
({\bf x_{0}+r},\tau_{0}) \rightarrow ({\bf x_{0}+r},\tau_{0}+\tau)
\rightarrow ({\bf x_{0}},\tau_{0}+\tau) \rightarrow ({\bf
x_{0}},\tau_{0})$ which follow either two
sides of a rectangle between ${\bf x_{0}}$ and ${\bf x_{0}+r}$ or a
single-step `staircase' route, to estimate $W({\bf r},\tau)$. 
To suppress the excited state
contributions, 
a simple APE smearing technique\cite{alb87,tep86} was used on the space-like 
variables. In this technique an iterative smearing procedure is used to 
construct Wilson loop (and glueball) operators with a very high degree of 
overlap with the lowest-lying state. 
In our single-link smoothing procedure, we replace every space-like link 
variable by  
\begin{equation}
U_{i} \rightarrow P\left[\alpha U_{i} + 
\frac{(1-\alpha)}{2}\sum_{s}U_{s}\right]
\label{eqn15}
\end{equation}
where the sum over ``s" refers to the ``staples", or 3-link paths 
bracketing the given link on either side in the spatial plane, and P denotes a
projection onto the group U(1), achieved by renormalizing the 
magnitude to unity.  We used a smearing parameter $\alpha = 0.7$
and up to ten iterations of the smearing process.

To further reduce the statistical errors, the timelike 
Wilson loops were constructed from 
``thermally averaged" temporal links\cite{par83}. That is, 
the temporal links $U_t$
in each Wilson loop were replaced by their thermal averages
\begin{equation}
\bar{U_t} = \int dU U \exp(-S[U])/ \int dU \exp(-S[U])
\label{eqn16}
\end{equation}
where the integration is done over the one link only, and depends on the
neighbouring links. For the U(1) model, the result can easily be
computed in terms of Bessel functions involving the `staples' adjacent
to the link in question. This was done for all temporal links except
those adjacent to the spatial legs of the loop, which are not
`independent'\cite{par83}. The procedure has a dramatic effect in
reducing the statistical noise, by up to an order of
magnitude\cite{mor97}, worth a factor of 100 in Monte Carlo runtime.

\begin{figure}[!h]
\scalebox{0.45}{\includegraphics{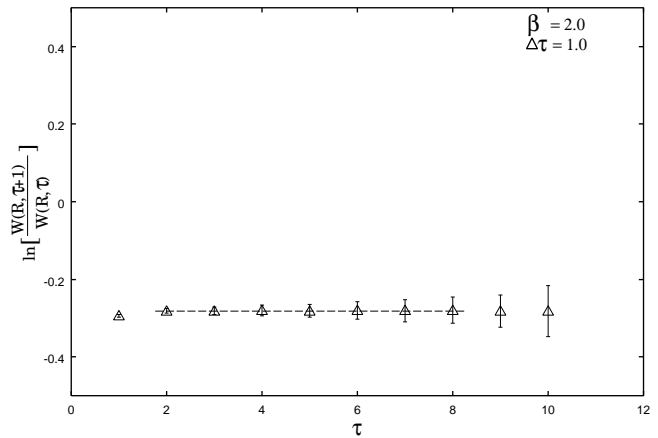}}
\caption{
\label{fig2}
Logarithmic ratio of Wilson loops as a function of $\tau$ 
for fixed $R=4$ 
at $\beta = 2.0$ and $\Delta \tau =1.0$. The dashed horizontal line 
indicates the plateau value.}
\end{figure}

The Wilson loop values $W({\bf r},\tau)$ are expected to decrease
exponentially with Euclidean time $\tau$. 
A typical plot of the logrithmic ratios of successive loop values
 is shown in Figure \ref{fig2}  for
$\beta = 2.0$, $\Delta \tau =1.0$ and $ R = \|{\bf r}\| =4$. 
It can be seen that with the heavy smearing we have used, a flat
`plateau' is attained virtually straight away.
The Wilson loops are therefore fitted  with the simple form
\begin{equation}
W({\bf r},\tau ) = a\mbox{e}^{-\tau V(\bf r)}
\label{eqn17}
\end{equation}
to determine the `effective' potential $V(\bf r)$. 

\subsection{Glueball masses}

Estimates for the glueball masses were obtained from the  time-like 
correlations between spatial Wilson loop operators $\Phi_{i}(\tau))$,
\begin{equation}
C(\tau) = \langle \Phi_i(\tau)^{\dagger}\Phi_i(\tau) \rangle
\label{eqn18}
\end{equation}
 in a standard fashion.
 As the temporal separation becomes large, the above correlator tends to be
dominated by the lowest energy state carrying the quantum numbers of 
$\Phi$.
If these quantum numbers coincide with those of the vacuum state, one then
looks at the next higher energy state. So before taking  the large
Euclidean time
limit, one subtracts the vacuum contribution from the correlator.
Thus 
\begin{equation}
\bar{\Phi}_{i}(\tau ) = \Phi_{i}(\tau ) - <0|\Phi_{i}(\tau )|0>
\label{eqn19}
\end{equation}
is a gauge invariant, translationally invariant, vacuum-subtracted operator 
capable of creating a glueball out of the vacuum.  
As a function of the temporal separation $\tau $, and with periodic
boundary conditions, the 
 correlation function is expected to behave as
\begin{eqnarray}
 C(\tau ) & = & < \bar{\Phi}^{\dagger}_{i}(\tau)\bar{\Phi}_{i}(0)>
 \nonumber \\
 & \simeq &  
  c_{1}(\exp(-m_{i}(\tau)) + \exp(-m_{i}(T-\tau))) \nonumber \\
 & & + \mbox{(excited state contributions)}
 \label{eqn21}
\end{eqnarray}
where $m_{i}$ is the mass of the lowest glueball state in that sector,
 and $T=N_{\tau}a_t$ is the extent of the periodic lattice in 
the time direction.
We project out states with momentum ${\bf k} =0$ and spin $J=0$ by 
summing over all lattice translations and rotations of the
 operators involved in $\Phi_{i}$.
In the present case, we study only the lowest-lying `antisymmetric' (PC
= - -) and `symmetric' (PC = ++) glueball states, corresponding to
operators $\Phi_{i}$ which are the sine and cosine, respectively, of the
sum of the link angles around the Wilson loop in question.

The statistical fluctuations in $C(\tau )$  
are given by\cite{bra90} 
\begin{equation}
\sigma \rightarrow \frac{C(0)}{\sqrt{N}}
\label{eqn22}
\end{equation}
Thus, the signal-to-noise ratio collapses as  
$C(\tau)$ falls exponentially fast with $\tau$. 
Hence, it becomes 
important to use a glueball operator for which the overlap with the glueball 
state of interest is strong for small lattice spacing, and such that  
$C(\tau )$ 
attains 
its asymptotic form as quickly as possible. For such an operator, the 
signal-to-noise ratio is also optimal\cite{bra90}. Such operators 
can be constructed by 
exploiting link smearing and variational techniques\cite{mor97,tep99}.

In the strong coupling limit  $\beta = 0$, the 
plaquette operator
$U_{P}$ will create a symmetric glueball state from  the 
vacuum. For large $\beta$, however, the glueball wave functions are 
expected to spread out and become more diffuse. To obtain a good overlap 
with the ground state in each sector at weak coupling, we need large, 
smooth operators $\phi_{i}$ on the lattice scale. An optimized operator is 
found by a variational technique, following Morningstar and Peardon\cite{mor97} 
and Teper\cite{tep99}. First, we calculate the correlation functions for square 
$n\times n$ Wilson loops with $m$ smearings, and determine the values of 
$m$ and $n$ for which the ratio $(C(1)/C(0))_{nm}$ is a maximum. In the 
second pass, an optimized glueball operator was found as a linear 
combination of the basic operators $\phi_{i}$,
\begin{equation}
\Phi (\tau) = \sum_{\alpha}v_{i\alpha}\phi_{i\alpha}(\tau)
\label{eqn23}
\end{equation}
where the index $\alpha$ runs over the rectangular Wilson loops with 
dimensions $l_{x}=[n-1, n+1]$, $l_{y} =[n-1, n+1]$ and smearings 
$n_{s}=[m-1, m+1]$ with $n$ and $m$ as determined in the first pass, 
making 27 operators in all. The $27\times 27$ correlation matrix was 
computed
\begin{equation}
C_{i\alpha \beta}(\tau ) = \sum_{\tau 0}<0|\bar{\phi}_{i\alpha}(\tau 
+\tau_{0})\bar{\phi}_{i\beta}(\tau_{0})|>
\label{eqn24}
\end{equation}
where $\bar{\phi}_{i\alpha}(\tau)$ is a vacuum-subtracted operator
\begin{displaymath}
\bar{\phi}_{i\alpha}(\tau)= \phi_{i\alpha}(\tau) 
-<0|\phi_{i\alpha}(\tau)|0>
\label{eqn25}
\end{displaymath}
The coefficients $v_{i\alpha}$ were then determined by minimizing the 
effective mass at $\tau =1$
\begin{equation}
\tilde{m}(1) = -\frac{1}{a_t}\ln\left[\frac{\sum_{\alpha \beta}v_{i\alpha}
v_{i\beta}
C_{i\alpha \beta}(1)}{\sum_{\alpha \beta}v_{i\alpha}v_{i\beta}
C_{i\alpha \beta}(0)}\right]
\label{eqn26}
\end{equation}
Let ${\bf{v}}_{i}$ denote a column vector whose elements are the optimal 
values of the coefficients $v_{i\alpha}$,
then the column vector ${\bf{v}}_{i}$ formed from the $v_{i\alpha}$ is 
the solution of an eigenvalue equation
\begin{equation}
C(1){\bf{v}}_{i} = \mbox{e}^{-a_t\tilde{m}(1)}C(0){\bf{v}}_{i}
\label{eqn27}
\end{equation} 
The eigenvector ${\bf{v}}_{0}$ corresponding to the largest eigenvalue $
\mbox{e}^{-a_t\tilde{m}}$ then yields the coefficients  for the operator
$\Phi_{i}(\tau )$ which best overlaps the lowest-lying state.

\begin{figure}[!h]
\scalebox{0.45}{\includegraphics{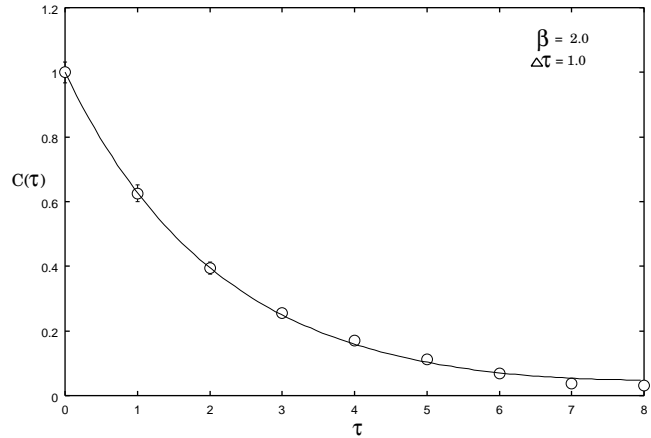}}
\caption{
\label{fig3}
Antisymmetric 
glueball correlation function $C(\tau)$ for the $0^{--}$ channel against 
$\tau$ 
at $\Delta 
\tau =1.0$ and $\beta =2.0$. The solid curve is a fit to the 
simulation results using   
Eq. (\ref{eqn29}).}
\end{figure}

A third pass was made to estimate the optimized correlation 
function
\begin{equation}
C_{i}(\tau ) = \sum_{\tau_0}<0|\bar{\Phi}_{i}(\tau + \tau_{0})
\bar{\Phi}_{i}(\tau_{0})|0>
\label{eqn28}
\end{equation}
Finally the optimized correlation function was fitted with the simple form 
\begin{equation}
C_{i}(\tau ) = c_{1}\cosh m_{i}(\frac{T}{2}-\tau )
\label{eqn29}
\end{equation}
to determine the glueball mass estimates.
Figure \ref{fig3} shows an example of the correlation function and fit for the antisymmetric state at
$\beta = 2$ and $\Dtau = 1$.
It can be seen that the form (\ref{eqn29}) fits the data very well.

\section{Simulation results at finite lattice size}

Simulations were carried out for lattices of $ N_s^{2}\times N_{t}$ sites, 
with $N_s=16$ and $N_{t}$ ranging from 16 to 48 
 sites, with periodic boundary conditions. Each 
run involved 250,000  
sweeps (350,000 for high anisotropy) of the lattice, with 50,000 sweeps 
(100,000 high anisotropy) discarded to
 allow for equilibrum, and configurations recorded every 250  sweeps 
thereafter. Coupling values from $\beta = 1.0$ to
 3.0 were explored at anisotropies $\Delta \tau$ 
ranging from 1 to 1/3. We fixed $\Dtau =16/N_t$ in the first pass, so that the 
lattice size
remains fixed at $16a_s$ in all directions.
At strong couplings (small $\beta$), we expect the behaviour to be the
same as in the bulk system, but at weaker couplings (large $\beta$) the
finite-size/finite-temperature corrections will become more important.
We shall monitor our data for signs of these effects. The results for the 
Euclidean case $\Dtau =1$  are listed in Table \ref{tabeucl}.

\begin{table}[!h]
\caption{
\label{tabeucl}
Monte Carlo estimates for the mean plaquette $<P>$, string 
tension $K$, 
symmetric and antisymmetric glueball masses $M_{0^{++}}$, $M_{0^{--}}$, 
and the mass ratio $R_{M}$ in the
Euclidean case  $\Delta\tau =1.0$.}
\begin{ruledtabular}
\begin{tabular}{cccccc}
 $\beta$ & $<P>$ & K & $M_{0^{++}}$ & $M_{0^{--}}$ & $R_{M}$ \\ \hline 
1.0 & 0.475 & 0.674 & 2.69(1) & 2.6(1) & 1.00(6)\\
1.35 & 0.629 &0.343(6) & 2.14(5)&1.65(5) & 1.29(5)\\
1.41 & 0.656 &0.286(5) & 2.08(5)&1.58(3) &1.31(7)\\
1.55 & 0.704 &0.200(1) &1.79(5)&1.26(3) & 1.41(7)\\
1.70 & 0.748  &0.122 & 1.41(3) & 0.88(1) & 1.60(8)\\
1.90 & 0.790 & 0.082 & 1.14(3) & 0.54(1) & 2.0(1)\\
2.0 & 0.806 & 0.050 & 0.79(1) & 0.44(1) & 1.78(9)\\
2.25 & 0.834 & 0.022 & 0.50(2) & 0.236(9) & 2.1(1)\\
2.5 & 0.854 & 0.012  & 0.34(2) & 0.165(9) & 2.0(1)\\
2.75 & 0.869 & 0.009 &  &  &    \\
3.0  & 0.881 & 0.010 &  &  &     \\
\end{tabular}
\end{ruledtabular}
\end{table}

\subsection{Mean Plaquette}

\begin{figure}[!h]
\scalebox{0.45}{\includegraphics{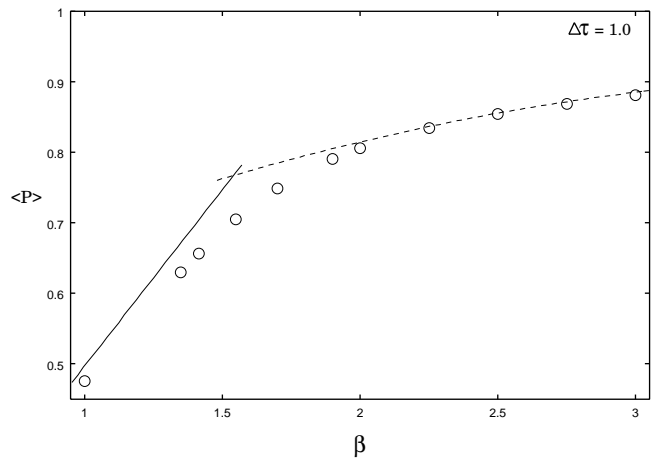}}
\caption{
\label{fig4}
The mean plaquette 
as a function of $\beta$ at $\Delta \tau =1$. The circles are our Monte  
Carlo estimates.
The solid curve represents the $O(\beta )$ 
strong-coupling expansion \protect\cite{bha80}  
and the dashed curve represents the 
$O(1/\beta^{4})$ weak-coupling expansion \protect\cite{hor81}.}
\end{figure}

Figure \ref{fig4} shows the behaviour of the mean spatial plaquette $<P>$ 
for different 
$\beta $, at fixed $\Dtau=1$ (Euclidean, isotropic case). 
A strong coupling perturbation series expansion has been obtained for
this quantity by Bhanot and Creutz\cite{bha80} to order $\beta$, and 
a weak coupling series to order $1/\beta^5$ by Horsley and
Wolff\cite{hor81}. These series are represented by solid and dashed lines on
the graph, respectively. 
It can be seen that the data follow the strong-coupling expansion for
$\beta \leq 1.5$, and match the weak-coupling expansion quite closely
beyond $\beta \simeq 2$.
The variation of $<P>$ with coupling is extremely 
smooth, with no sign of any phase transition, as we should expect. The 
cross-over from strong to weak coupling seems 
to take place quite rapidly in the region
$\beta \approx 1.8 - 2.0$.
 Horsley and Wolff \cite{hor81} 
investigated the effect of a finite-size lattice in their 
weak-coupling expansion calculations. They found that such effects enter at 
order $1/\beta^{2}$ as a correction of order $1/L^{D}$, where $L$ is the 
lattice size and $D$ is the number of dimensions.  Thus for the compact U(1) 
model in 3-dimensions 
and  $L>10$, the finite size effects should be essentially 
negligible.

\begin{figure}[!h]
\scalebox{0.45}{\includegraphics{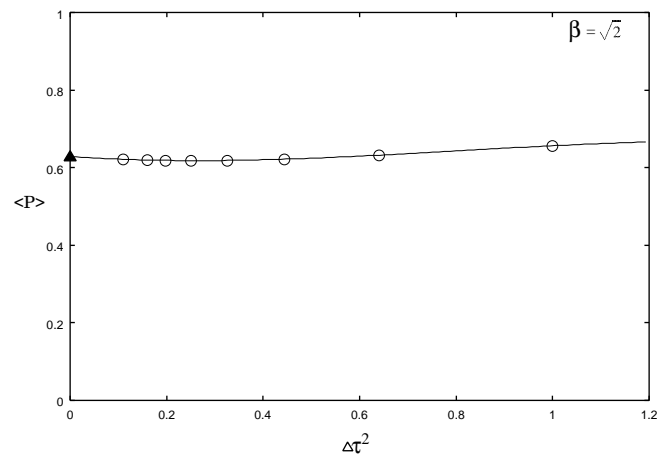}}
\caption{
\label{fig5}
The mean plaquette  as a function of $\Delta \tau^{2}$ at
$\beta =\sqrt{2}$. The solid curve is a cubic fit to the data,  
extrapolated to the Hamiltonian limit. The triangle shows the Hamiltonian
 series estimate \protect\cite{ham00} in that limit.}
\end{figure}

Figure \ref{fig5} shows a plot of our estimates
of $<P>$ as a function of anisotropy \cite{irv84}\footnote{The physical anisotropy, 
as measured spatial versus temporal correlation lengths for instance, will differ from the
`bare' anisotropy $\Dtau$. Since we are only interested in the Hamiltonian limit 
$\Dtau \rightarrow 0$, however, this is of no concern for our present purposes}
${\Delta \tau}^{2}$ for the case
$\beta = \sqrt{2}$. 
It can be seen that $<P>$ remains almost constant.
We would like to make contact with previous Hamiltonian studies
by showing that the mean plaquette value approaches
to previously known values in the Hamiltonian limit
 $\Delta \tau \rightarrow 0$. 
The extrapolation
was performed using a simple cubic fit in powers of $\Dtau^{2}$.  
In this limit our results agree very well with the
Hamiltonian estimate obtained by Hamer et al\cite{ham00}.
Our general estimates in the Hamiltonian limit $\Dtau =0$ are listed in 
Table \ref{tabhiml}.

\begin{table}[!h]
\caption{
\label{tabhiml}
Monte Carlo estimates for the mean plaquette $<P>$, string
tension $K$,
symmetric and antisymmetric glueball masses $M_{0^{++}}$, $M_{0^{--}}$,
and the mass ratio $R_{M}$ in the Hamiltonian limit  $\Delta\tau =0$.}
\begin{ruledtabular}
\begin{tabular}{cccccc}
$\beta$ & $<P>$ & K & $M_{0^{++}}$ & $M_{0^{--}}$ & $R_{M}$  \\ \hline
1.0 & 0.397(6) & 0.302(3) & 1.9(1) & 1.45(7) & 1.32(1) \\
1.35 & 0.599(2) & 0.132(8) & 1.4(1) & 0.9(1) & 1.5(2)\\
1.41 & 0.628(1) & 0.104(7) & 1.05(6) &  0.60(7) & 1.7(2)\\
1.55 & 0.651(1) & 0.085(9) &0.8(2) & 0.3(1) & 2.1(1.1)\\
1.70 & 0.695(3) & 0.021(6) & 0.5(3) & 0.24(9) & 2.2(1.5)\\
1.90 & 0.716(4) & 0.018(7) &0.3(1) & 0.17(9) & 2.2(1.5)\\
2.0 & 0.740(4) & 0.015(1) & 0.2(1) & 0.10(6) &   \\
2.25 & 0.779(4) & 0.008(4) & 0.15(9) & 0.07(5) &    \\
2.5 & 0.796(5) & 0.005(8) &  & &   \\
2.75 & 0.820(5) & & & &  \\
3.0  & 0.839(4) & & & & \\
\end{tabular}
\end{ruledtabular}
\end{table}

\begin{figure}[!h]
\scalebox{0.45}{\includegraphics{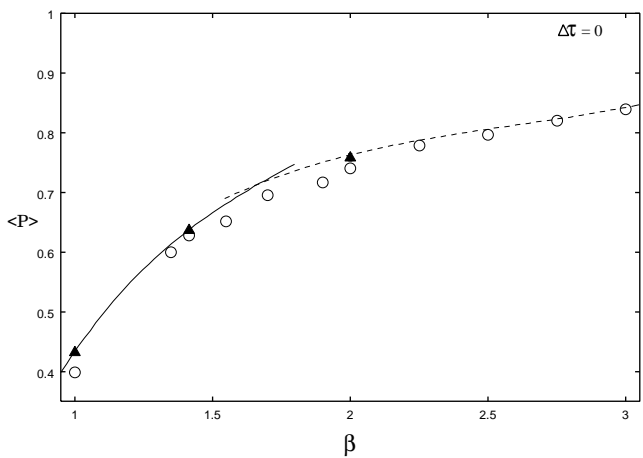}}
\caption{
\label{fig6}
Mean plaquette estimates as a function of  
 $\beta$ at $\Delta \tau =0$. 
 Our present Monte Carlo estimates are shown by circles.
 The strong-coupling \protect\cite{ham92}
 and weak-coupling \protect\cite{ham93} series predictions are shown as solid 
 and dashed lines respectively. The
 GFMC estimates \protect\cite{ham00} are represented by 
the solid triangles.} 
\end{figure}

Figure \ref{fig6} graphs the resulting estimates of $<P>$ in the
Hamiltonian limit $\Delta \tau = 0$ as a function of
coupling $\beta$. The weak-coupling\cite{ham93} and 
strong-coupling\cite{ham92} series
predictions are shown as dashed and solid lines on the graph
respectively, while some previous Greens Function Monte Carlo estimates
\cite{ham00} are shown as triangles. Our present results are generally
in reasonable agreement with the earlier ones, if perhaps a little low
in places. It can be seen that the crossover from strong to weak
coupling behaviour again occurs at around $\beta \simeq 1.8$.

\subsection{Static quark potential and string tension}

\begin{figure}[!h]
\scalebox{0.45}{\includegraphics{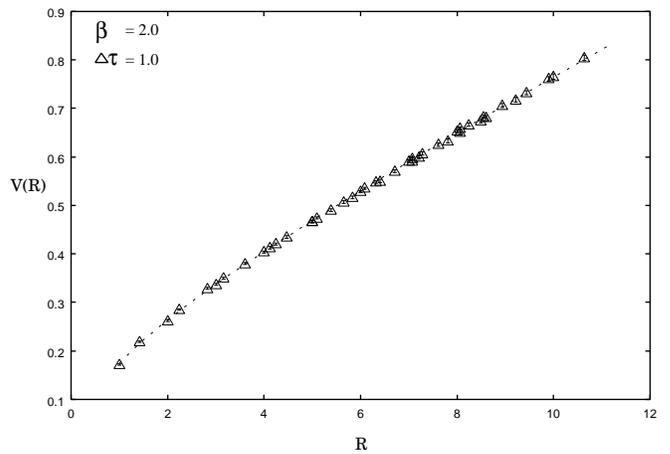}}
\caption{
\label{fig7}
The static-quark potential $V(R)$ as a function of  
the separation R. 
This plot involves measurements at $\beta=2.0$ for $\Delta \tau
= 1.0$ with 10 smearing sweeps  at smearing parameter $\alpha = 0.7$.
The  errors are smaller than the  symbols. The dashed 
line is a fit to the form $V(R)=a+bR+c\mbox{ln}(R)$.}
\end{figure}

Figure \ref{fig7} shows a  graph of the static quark potential V(R)   as a 
function of radius R at $\beta = 2.0$ and $\Delta \tau = 1.0$.
To extract the string tension, the curve is fitted by
a form
  \begin{equation}
  V(R) = a +b\ln R+\sigma R,
\label{eqn29a}
  \end{equation}
including a logarithmic Coulomb term as expected for classical 
QED in (2+1) dimensions which dominates the behaviour at small distances, 
and
a linear term as predicted by Polyakov\cite{pol78} and G{\" o}pfert and 
Mack\cite{gop82} dominating the behaviour at large distances. The linear
behaviour at large distances is very clear, but the data do not extend
to very small distances, so there is no real test of the presumed
logarithmic behaviour in this regime.

\begin{figure}[!h]
\scalebox{0.45}{\includegraphics{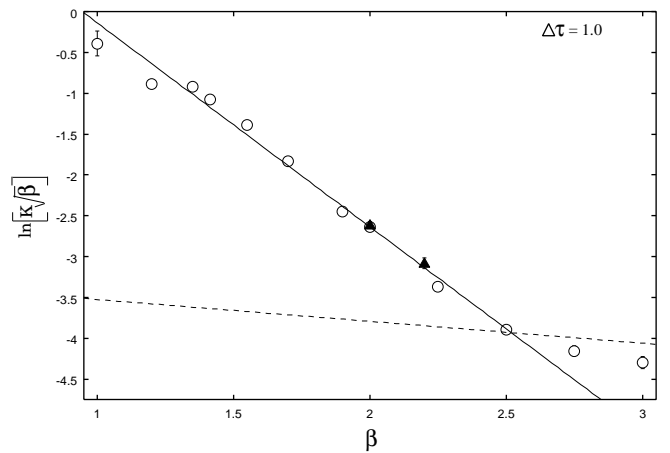}}
\caption{
\label{fig8}
$\mbox{ln}(K\sqrt{\beta})$ as a function of $\beta$
at $\Dtau=1.0$. The solid curve 
represents the  predicted asymptotic form, eq. (\protect\ref{eqn6}), 
with our estimated value for the
normalization constant, $c=44.0\pm 0.4$. 
The dashed line represents the finite size scaling
behaviour\protect\cite{ham93}. The solid triangles show the previous estimates 
of Irb{\" a}ck and Peterson \protect\cite{irb87}. } 
\end{figure}

Figure \ref{fig8} shows the behaviour of the fitted value of the string 
tension  $K = \sigma a^2$ 
as a function of $\beta$ for the Euclidean, isotropic case 
($\Delta \tau =1$). The solid line on the graph represents the form
(\ref{eqn6}) predicted by G{\" o}pfert and Mack\cite{gop82}, using a
value of $c = 44.0\pm 0.4$. It can be seen that this form represents
the data rather well over a range $1.41 \leq \beta \leq 2.5$;
in fact an unconstrainted fit to the data gives a slope of 
$2.49\pm 0.15$, extremely 
close to the predicted value 2.494. The 
coefficient $c$, however, 
 is much bigger than the value $c = 8$ predicted in the
classical approximation. It would be interesting to explore how
higher-order quantum corrections affect the prediction for $c$.

The dashed line in Figure \ref{fig8} gives some idea of the 
expected finite-size
scaling corrections to the string tension. These have not been
calculated explicitly for the Euclidean model, as far as we are aware,
but in the Hamiltonian version the string tension at weak coupling is
found\cite{ham93} to behave as
\begin{equation}
K = \frac{1}{2\beta L}
\label{eqn30}
\end{equation}
where $L = N_s$ is the lattice size (here $L = 16$), and this is
represented by the dashed line. This would predict that the string
tension will be dominated by finite-size corrections beyond $\beta
\simeq 2.5$, and indeed the Monte Carlo estimates do flatten out beyond
that point, although at a level below equation (\ref{eqn30}). 

\begin{figure}[!h]
\scalebox{0.45}{\includegraphics{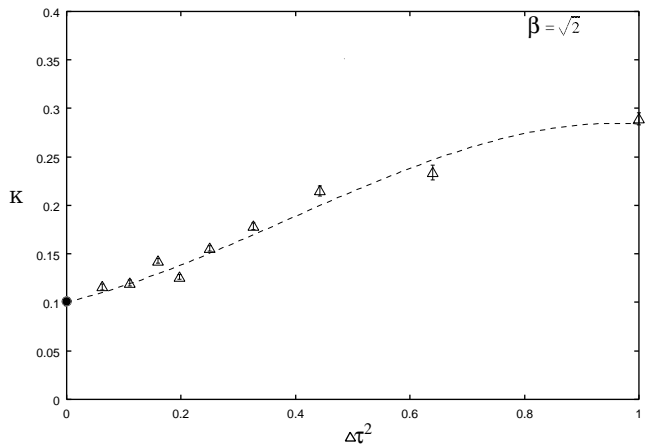}}
\caption{
\label{fig9}
String tension, K,  as a function of $\Delta \tau^{2}$ at 
$\beta =\sqrt{2}$. 
An extrapolation to the Hamiltonian limit is performed by a cubic fit and 
shown by 
the dashed line. Our MC estimates are shown by triangles and  
an earlier 
Hamiltonian series estimate \protect\cite{ham92} is  shown by a 
solid circle.}
\end{figure}

Figure \ref{fig9} shows the behaviour of the string tension $K$
as a function of the anisotropy ${\Delta \tau}^{2}$, for fixed
coupling $\beta = \sqrt{2}$.
An extrapolation to the Hamiltonian limit $\Dtau \rightarrow 0$ is
performed by a simple cubic fit. Again the extrapolation agrees 
well with earlier Hamiltonian estimates\cite{ham94}. Note that this
quantity depends rather strongly on $\Dtau$: there is a factor of three
difference between the values at $\Dtau = 0$ and $\Dtau = 1$.
Extrapolating to $\Dtau \rightarrow 0$, estimates of the string tension in 
the Hamiltonian limit are obtained for various $\beta$ values. 

\begin{figure}[!h]
\scalebox{0.45}{\includegraphics{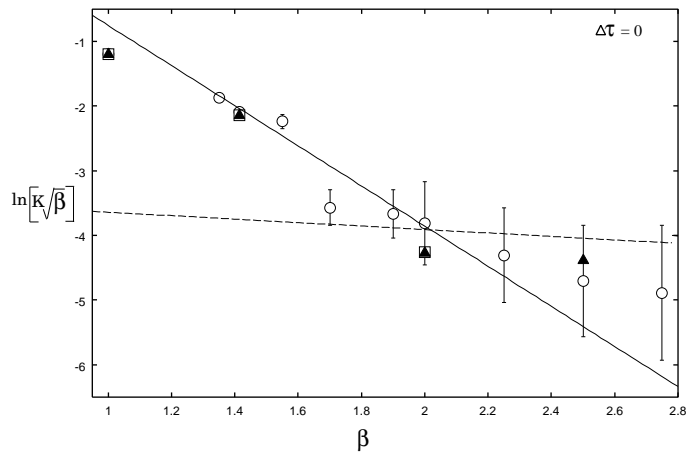}}
\caption{
\label{fig10}
Graph  showing  estimates of the string tension in the Hamiltonian 
limit
as a function of  $\beta$. The solid line is a least square fit to the form
$K=\beta^{-1/2}\mbox{exp}(a_{0}-a_{1}\beta )$, with 
$a_{0} = 2.359$ and $a_{1}=3.159$. 
 The dashed line represents the 
finite size scaling behaviour\protect\cite{ham93}. 
Earlier results from an exact linked cluster expansion\protect\cite{irv84} and
quantum Monte Carlo simulations \protect\cite{ham94} are shown as solid
triangles and squares respectively.}
\end{figure}

Our estimates of the string tension in the Hamiltonian limit are
graphed in Figure \ref{fig10}, together with
earlier results from 
an `exact linked cluster expansion'\cite{irv84} and a quantum Monte
Carlo simulation\cite{ham94}.
It can be seen that our values are consistent with earlier results,
though less accurate, and extend further into the weak-coupling region.
The solid line in the graph represents a least-square fit of the 
weak-coupling asymptotic form
(\ref{eqn6}), with $v(0)=0.3214$ and $c=46.7\pm 0.4$. This form
represents the data well for $1.35 \leq \beta \leq 2.0$. Beyond $\beta
=2$ the string tension is consistent, within errors, with the
finite-size behaviour predicted by equation (\ref{eqn30}), which is shown as
a dashed line.

\subsection{Glueball Masses}

\begin{figure}[!h]
\scalebox{0.45}{\includegraphics{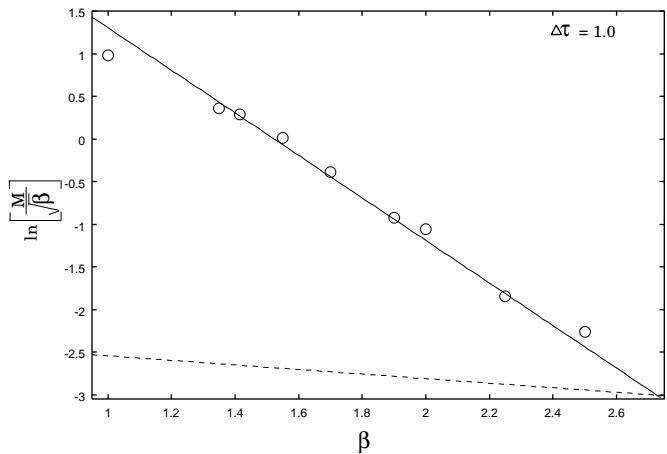}}
\caption{
\label{fig11}
The scaling behaviour of the antisymmetric mass gap against $\beta$ 
at $\Delta \tau = 
1.0$. The solid line is a fit of the form eq. (\protect\ref{eqn31}). The errors
are smaller than the symbols.
The dashed line  shows the finite size scaling behaviour 
\protect\cite{wei99}.}
\end{figure}

Figure \ref{fig11} shows results for the antisymmetric $0^{--}$ glueball
mass against $\beta$ for the isotropic Euclidean case $\Delta \tau = 1$.
The solid line on the graph is a fit 
to the data over the range 
$1.4\leq \beta \leq 2.25$  using the predicted asymptotic
form, equation (\ref{eqn5}), but with an additional multiplying
constant:
\begin{equation}
M = am_D = c_1\sqrt{8\pi^2\beta}\exp(-\pi^2\beta v(0))
\label{eqn31}
\end{equation}
where $c_1 = 5.23\pm 0.11$ when adjusted to fit the data. Thus the slope,
$2.48\pm 0.09$, 
 of
the data matches the predicted asymptotic form very nicely, but the
coefficient is too large by a factor of $5.2$. It would again be
interesting to explore whether this discrepancy could be due to quantum
corrections.

The expected finite-size scaling behaviour of the mass gap near the
continuum critical point in this model
is not known; but Weigel and Janke\cite{wei99} have performed a Monte
Carlo simulation for an O(2) spin model in three dimensions which should
lie in the same universality class, obtaining
 
\begin{equation}
M \sim 1.3218/L 
\label{eqn32}
\end{equation}
for the magnetic gap. The dashed line in Figure \ref{fig11} shows this
prediction for $L=16$. It can be seen that the Euclidean mass gap should
not be affected by finite-size corrections until $\beta \geq 2.8$.

\begin{figure}[!h]
\scalebox{0.45}{\includegraphics{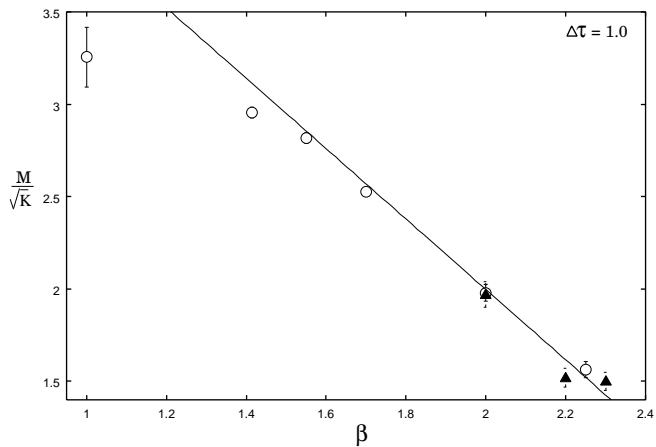}}
\caption{
\label{fig12}
The dimensionless ratio $M/\sqrt{K}$ as a function 
$\beta$. Our estimates are shown by circles and solid triangles  
 show the earlier results of Teper 
\protect\cite{tep99}. The solid curve 
represents the  predicted weak-coupling behaviour.}
\end{figure}

To check the consistency of our method, we plot the dimensionless ratio
of the antisymmetric mass gap over the square root of the string tension
against $\beta$ together with the results of Teper\cite{tep99} in Figure
\ref{fig12}. The agreement is excellent. The solid line gives the ratio
of the fits in Figures \ref{fig8} and \ref{fig11}, and shows how this
ratio vanishes exponentially in the weak-coupling limit, whereas in
four-dimensional confining theories it goes to a constant.

\begin{figure}[!h]
\scalebox{0.45}{\includegraphics{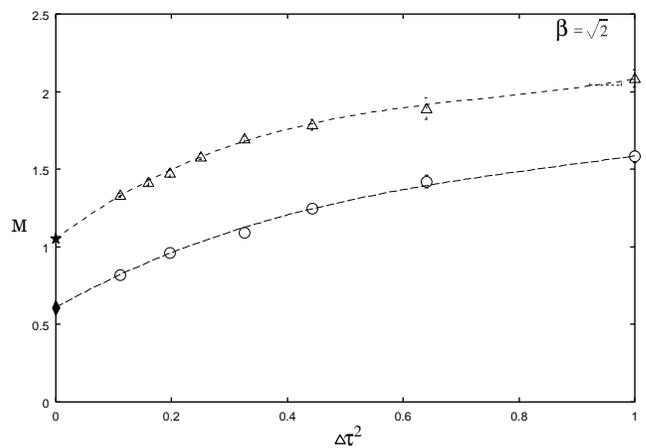}}
\caption{
\label{fig13}
Estimates of the masses of $0^{++}$ and $0^{--}$ glueballs  
against $\Delta \tau^{2}$. Results at $\beta = 
\sqrt{2}$  for the $0^{++}$ and $0^{--}$ are labeled by 
circles and triangle respectively.
The solid and dashed curves are the cubic fits to the data  extrapolated to 
the Hamiltonian limit. The  
series estimates of Hamer et al \protect\cite{ham92} in the limit 
$\Dtau \rightarrow 0$, for symmetric and antisymmetric channels are 
shown as a star and diamond respectively.}
\end{figure}

Figure \ref{fig13} shows the behaviour of the glueball masses as functions 
of ${\Delta
\tau}^{2}$ for $\beta = \sqrt{2}$. The extrapolation to the Hamiltonian 
limit is performed by a simple cubic fit in powers of $\Delta \tau^2$.
In this limit we 
reproduce the earlier estimates of Hamer et al\cite{ham92} for the $0^{--}$
and $0^{++}$ states. 

\begin{figure}[!h]
\scalebox{0.45}{\includegraphics{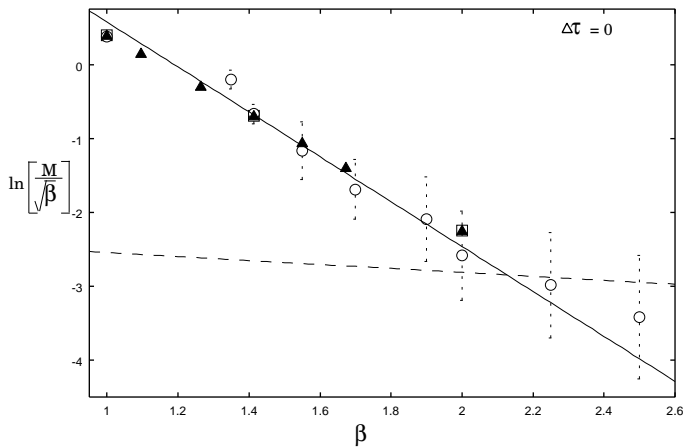}}
\caption{
\label{fig14}
Hamiltonian estimates  of the  antisymmetric mass gap 
plotted 
as a function of $\beta$. The 
solid curve is the fit to the data for $1.4<\beta <2.25$. The dashed line 
represents the finite size effects \protect\cite{wei99}. The previous
results from series expansion \protect\cite{ham92} and quantum Monte Carlo
calculations \protect\cite{ham94} are shown as solid triangles and open squares
respectively.}
\end{figure}

Estimates of the antisymmetric mass gap in the Hamiltonian limit $\Delta
\tau =0$ are graphed against $\beta$ in Figure \ref{fig14}. Also shown
are results from previous strong-coupling series extrapolations\cite{ham92}
and quantum Monte Carlo calculations\cite{ham94}. It can be seen that our
present results agree with previous estimates but are less accurate.
The solid line is a fit to our data for $1.4 \leq \beta \leq 2.25$ of 
the form (\ref{eqn29}), with $v(0) = 0.3214$
and $c_1 = 5.5\pm 0.2$, which is similar to the coefficient found in 
the
Euclidean case.
The fit  to the data gives a  slope of $3.1\pm 0.2$ and an 
intercept of $3.6\pm 0.3$ of the scaling curve.
We note that the exponential slope in
previous studies is generally somewhat less than this, as tabled in 
\cite{john97} and as illustrated by the black triangles in Figure 
\ref{fig14}. 
The dashed line represents the finite-size scaling behaviour, equation
(\ref{eqn32}), which we assume holds in the Hamiltonian limit also, for
want of better information. It can be seen that the finite-size
corrections are predicted to dominate for $\beta \geq 2.2$, but the date
are not accurate enough at weak couplings to establish whether this is 
really the case.

\begin{figure}[!h]
\scalebox{0.45}{\includegraphics{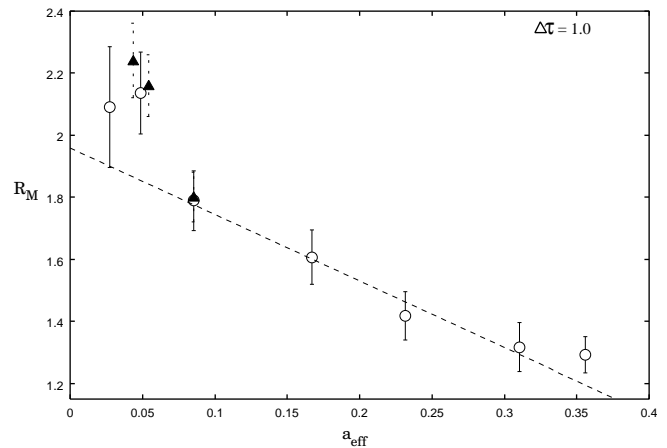}}
\caption{
\label{fig15}
A graph showing estimates of the mass ratio $R_M$ as a function of the
effective spacing, $a_{eff}$, at $\Delta \tau =1.0$. Our present estimates are 
shown by the circles. The dashed line is a linear fit to the data over the 
range 
$0.08\leq a_{eff} \leq 0.31$. The solid triangles show the previous estimates 
of Teper \protect\cite{tep99}.} 
\end{figure}

Finally, 
Figure \ref{fig15} displays the behaviour of the dimensionless mass
ratio,
\begin{equation}
R_M = M(0^{++})/M(0^{--})
\label{eqn33}
\end{equation}
for the Euclidean case $\Delta \tau = 1$. As in the (3+1)D confining
theories, we may expect that quantities of this sort will approach their
weak-coupling or continuum limits with corrections of $O(1/a_{eff})$,
where $a_{eff}$ is the effective lattice spacing in `physical' units
when the mass gap has been renormalized to a constant. Hence for our
present purposes we {\it define} $a_{eff}$ from equation (\ref{eqn5})
as
\begin{equation}
a_{eff} = \sqrt{8\pi^2\beta}\exp(-\pi^2\beta v(0))
\label{eqn34}
\end{equation}
with $v(0) = 0.2527$ for the Euclidean case. The mass ratio is plotted
against $a_{eff}$ in Figure \ref{fig15}. 
At weak coupling, we expect the theory to 
approach a theory of free bosons\cite{gop82} so that the symmetric state will 
be 
composed of two $0^{--}$ bosons and the mass ratio should approach 
two. Our results show that as
$a_{eff}$ goes to zero, the mass ratio rises to around the expected value of
$2.0$. A
linear fit to the data from $0.08 \leq a_{eff} \leq 0.32$ gives an 
intercept 
$R_M = 1.95\pm 0.05$.
However, we note that the last two of our estimates, together with two
from Teper\cite{tep99}, lie considerably {\it above} $R_M = 2$. 
In the bulk system, of course, the ratio cannot rise above 2, because 
it is always possible to construct a $0^{++}$ state out of two $0^{--}$
mesons.
These points
correspond to couplings $\beta > 2$, and we conjecture that they may
be affected by finite-size corrections: a simulation on a larger lattice
would be necessary to check on this point. Our last two points have not
been included in the fit. 

\begin{figure}[!h]
\scalebox{0.45}{\includegraphics{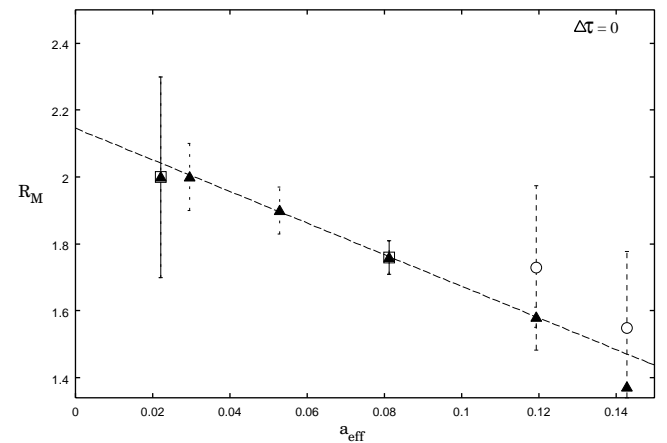}}
\caption{
\label{fig16}
Mass ratio in the Hamiltonian limit as a function of the
effective spacing, $a_{eff}$, at $\Delta \tau =0$. Our MC estimates are 
shown by the circles. 
Earlier series  \protect\cite{ham92} and quantum Monte Carlo 
\protect\cite{ham00a} results
are shown by solid triangles and open squares respectively. The dashed line is 
a linear fit to the earlier data from  series expansions \protect\cite{ham92}
 over the range
$0.02\leq a_{eff} \leq 0.12$.} 
\end{figure}

Figure \ref{fig16} shows a similar graph for the Hamiltonian limit,
$\Delta \tau = 0$. Within errors, our present results are consistent
with earlier series\cite{ham92} and quantum Monte Carlo\cite{ham94} estimates, 
but are much 
less accurate, and tend to lie consistently on the high side. A
linear fit to the earlier data from $0.02\leq a_{eff} \leq 0.12$
gives $R_M = 2.14\pm 0.01 $.

\section{Conclusions}  

In this paper, we have applied standard Euclidean path integral Monte
Carlo methods to the U(1) model in (2+1) dimensions on an anisotropic
lattice, and taken the anisotropic limit $\Delta \tau \rightarrow
0$ to obtain the Hamiltonian limit of the model.

We have obtained the first clear picture of the static quark potential in
this model, showing very clear evidence of the linear confining
behaviour at large distances predicted by Polyakov\cite{pol78}. There is also
a turnover at short distances consistent with a logarithmic Coulomb
behaviour in that regime.

In the isotropic or Euclidean case $\Delta \tau = 1$, the string tension
and mass gap display an exponential decrease at weak couplings which is
in excellent agreement with the behaviour predicted by
Polyakov\cite{pol78} and G{\" o}pfert and Mack\cite{gop82}. Both
quantities, however, are 5-6 times larger in magnitude than the theory
predicts. It would be interesting to calculate whether higher-order
quantum corrections can account for this discrepancy.

The dimensionless ratio $M/\sqrt{K}$ scales exponentially to zero in the
weak-coupling or continuum limit, as predicted by the theory. The mass
ratio of the two lowest glueball states scales against the effective
lattice spacing towards a value close to $2.0$, as expected for a theory
of free scalar bosons, apart from some anomalous results at large
$\beta$ which we have ascribed to finite-lattice effects.
 
In the anisotropic or Hamiltonian limit $\Delta \tau \rightarrow 0$, our
results are less accurate, because of the extrapolation needed to reach
this limit. Nevertheless, the results are generally in good agreement
with those obtained by other methods. Once again, an exponential
behaviour of the string tension and glueball masses can be demonstrated
at weak coupling.
The dimensionless mass ratio again scales to a value near $2.0$. Because
the exponential slope is steeper, finite-size effects seem to be
somewhat more important in the Hamiltonian regime than in the Euclidean
one.

Our major object in this study was to compare the Euclidean PIMC
approach to quantum Monte Carlo methods such as Green's Function Monte
Carlo (GFMC) for obtaining estimates in the Hamiltonian limit. The PIMC
approach suffers from the disadvantage that an extrapolation is
necessary to reach the limit $\Delta \tau \rightarrow 0$; while GFMC
suffers from the major disadvantage that it relies on a `trial wave
function'. In the event, we have obtained much better results using
PIMC. A clear and consistent picture of the string tension and glueball
masses was obtained at weak coupling. Using GFMC, on the other hand,
only qualitative estimates of the string tension were obtained, and the
glueball mass estimates were virtually worthless\cite{ham00}. No doubt there
are many `tricks of the trade', such as smearing and variational
techniques, which could be used to improve the GFMC results; but we
found previously\cite{ham00a} that there is an unacceptably strong dependence
on the trial wave function using that technique, especially for large
lattice size. The PIMC technique seems to offer a much more robust and
unbiased approach to Hamiltonian lattice gauge theories.
Of course, one must generally expect the Hamiltonian estimates to be
less accurate than the Euclidean ones because of the extra extrapolation
involved.

We note that the PIMC results are still less accurate than some older
quantum Monte Carlo results of Hamer, Wang and Price\cite{ham94}. The
latter were obtained using a strong-coupling representation, however,
and this approach has been found to fail for non-Abelian
models\cite{ham94} due to the occurrence of a `minus sign' problem, as
mentioned in the introduction.

\acknowledgments

This work was supported by the Australian 
Research Council. We are grateful to Mr. Tim Byrnes for help with some
of the calculations. We are also grateful for access to the
computing facilities of the Australian Centre for Advanced Computing and
Communications (ac3) and the Australian Partnership for Adsvanced
Computing (APAC).

\end{document}